\documentclass[11pt]{article}
\usepackage{epsfig}
\usepackage{latexsym}
\usepackage{threeparttable}
\usepackage{amssymb,amsmath}
\usepackage{cite}

\newcommand{\beq}{\begin{eqnarray}}
\newcommand{\eeq}{\end{eqnarray}}
\newcommand{\be}{\begin{eqnarray*}}
\newcommand{\ee}{\end{eqnarray*}}

\newcommand{\D}{{\cal D}}
\newcommand{\Pom}{{\hspace{ -0.1em}I\hspace{-0.25em}P}}

\newcommand{\vp}{{\gamma^*}}
\newcommand{\qqbar}{q \overline{q}}

\newcommand{\xP}{x_\Pom}
\newcommand{\FD}{F_{2 \D}^{(3)} }

\begin{document}
\title{Nuclear shadowing in Glauber-Gribov theory with $Q^2$-evolution}

\author{
\begin{tabular}{c}
N\'estor~Armesto,$^1$ \hspace{1.em} Alexei~ B.~Kaidalov,$^2$ \\ Carlos~A.~Salgado,$^1$ \hspace{0.5em} and \hspace{0.5em} Konrad~Tywoniuk$^1$ \\
\\
$^1$ Departamento de F\'isica de Part\'iculas and IGFAE, \\
Universidade de Santiago de Compostela, \\ 
15706 Santiago de Compostela, Galicia, Spain \\ \\
$^2$ Institute of Theoretical and Experimental Physics,\\
117259 Moscow, Russia
\end{tabular}
}
\date{\today}
\maketitle

\begin{abstract}
We consider deep inelastic scattering off nuclei in the Regge limit within the Glauber-Gribov model. Using unitarized parton distribution functions for the proton, we find sizeable shadowing effects on the nuclear total and longitudinal structure functions, $F_2^A$ and $F_L^A$, in the low-$x$ limit. Extending a fan-diagram analysis for the large-mass region of coherent diffraction off nuclei to high $Q^2$, we also find significant shadowing effects in this kinematical regime. Finally, we discuss shortcomings of our approach and possible extensions of the model to other kinematical regimes.
\end{abstract}
\hspace{25pt} {\small {\bf PACS:} 12.38.-t, 24.85.+p, 25.30.-c}

\section{Introduction}

The Regge limit of QCD ($s\rightarrow \infty$ at constant $Q^2$) is of great interest both from the perspective of high-energy collider experiments and inasmuch as it poses a challenge for the theoretical understanding of the hadronic wave-function. The rapid growth of cross sections observed at high $Q^2$ and fairly low $x$ is expected to slow down due to unitarity as $x \rightarrow 0$. Compared to the nucleon case, in $\vp A$ collisions these effects are enhanced by the nuclear thickness factor $\sim A^{1/3}$. While a fully consistent description of nucleus-nucleus collisions at high energies is yet to be found, the characteristic unitarity corrections are expected to emerge in the context of deep-inelastic scattering (DIS) off nuclei at high energies.

Nuclear modifications of the total $\vp A$ cross sections are seen to arise for all $x > 10^{-2}$ \cite{Arneodo:1992wf}. Nuclear parton distribution functions (PDFs) have been extracted at LO and NLO from the available data in \cite{deFlorian:2003qf,Hirai:2007sx,Eskola:2009uj}. The resulting PDFs differ significantly at low momentum scales due to different choices of functional forms and data samples made by the various groups, especially in the gluon sector, but gives rather consistent results at higher $Q^2$. Yet, the restricted kinematic access of present-day experiments give weak constraints on the low-$x$ region interesting for high-energy experiments and from the point of view of unitarity corrections.

The depletion of the nuclear structure function observed at $x < 0.1$, compared to the incoherent superposition of $A$ $\vp p$ cross sections, is usually called shadowing (for a recent review, see \cite{Armesto:2006ph}). As mentioned before, these effects are expected to occur at high energies due to the unitarization of the total cross section. In the infinite momentum frame (of the fast nucleus) the depletion is interpreted as destructive interference between gluons originating from different nucleons \cite{Gribov:1981ac}. Similarly, in a frame where the target nucleus is at rest, the same effects arise due to multiple scattering of the projectile.

In the latter context, there emerges a critical length scale related to a change of the underlying space-time picture of the collision. The coherence length (or life-time) of a given fluctuation of the incoming projectile is given by
\beq
l_C = \frac{1}{Q} \frac{E_{LAB}}{Q} \simeq \frac{1}{ 2 m_N x} \;,
\eeq
in the limit $2 m_N E_{LAB} = W^2 \gg Q^2$, where $x$ is the Bjorken variable. At low energies, where $l_C$ is of the order of the internucleon distance, the projectile undergoes incoherent multiple scattering off the target. Remarkably, all higher-order rescatterings cancel and the total $\vp A$ cross section is simply given as a superposition of $\vp p$ collisions. The critical value is reached when the coherence length becomes of the order of the nuclear radius. For $l_C > R_A$, i.e. at $x < 1 \big/ 2 m_N R_A$, the projectile scatters coherently off all constituents of the nucleus at some given impact parameter. Despite the non-local nature of the interactions, the total cross section can be written in the form of a multiple scattering series, now including corrections from higher-order rescattering diagrams which lead to an overall depletion of the total cross section.

The analysis of rescattering of the lowest order Fock-states can readily be performed within the framework of the dipole model \cite{Mueller:1989st,Nikolaev:1990ja,Nikolaev:1995xu}. The dynamics of the scattering is encoded in the dipole-target cross section. Whereas the $\vp p$ cross section can be analyzed to great detail due to the wealth of available experimental data, the additional impact parameter dependence arising naturally in the nuclear case is largely unconstrained. There are several propositions for such an {\it ansatz} in the literature \cite{Kowalski:2003hm, Iancu:2003ge, Kowalski:2006hc}. An additional challenge to this type of models is to properly take into account scattering of higher-order Fock-states which dictate the strength of the rescattering and could, thus, underestimate nuclear shadowing. This problem was addressed in \cite{Kopeliovich:2008ek} where the propagation of a given massive dipole Fock-state in a large nucleus was modeled and shadowing from $\qqbar$+N$g$ configurations was included.

The large probability of rescattering and, thus, a large probability of diffraction dissociation, is known to arise from asymmetric dipole pairs and large dipole fluctuations. These involve a strong soft contribution already at leading twist in the cross section. A suitable framework, although not rigorously established within QCD, to treat these configurations is provided by the reggeon calculus \cite{Gribov:1968fc}, where rescattering of the projectile wave function is accounted for by including the full mass-spectrum of hadronic intermediate states. This is the so-called Gribov inelastic shadowing \cite{Gribov:1968jf,Gribov:1968gs} (for a comprehensive review see e.g. \cite{Capella:1999kv}). 

In a previous paper \cite{Armesto:2010ee}, we have analyzed the total, longitudinal and diffractive $\vp p$ cross sections in an unitarized model \cite{Capella:2000pe,Capella:2000hq} with $Q^2$-evolution at leading order (LO) in the strong coupling constant. The non-perturbative model \cite{Capella:2000pe,Capella:2000hq} included unitarity corrections by summing all multi-pomeron exchanges, thus providing initial conditions for QCD evolution equations down to $x \sim 10^{-8}$. The aim of the present work is to extend this analysis to DIS off nuclei. This will be done in the framework of the Glauber model \cite{Glauber:1959} with its extension to the relativistic case due to Gribov \cite{Gribov:1968jf,Gribov:1968gs}. This formalism relates the inclusive and diffractive $\vp p$ cross sections to the corresponding $\vp A$ ones by means of the AGK cutting rules \cite{Abramovsky:1973fm}.  The original, non-perturbative model \cite{Capella:2000pe,Capella:2000hq} was thus used to predict nuclear shadowing at low-$Q^2$ in good agreement with existing experimental data \cite{Armesto:2003fi}. 

Shadowing from Glauber-Gribov theory has also been calculated previously in a similar model based on parameterizations of existing data on inclusive and diffractive cross sections \cite{Frankfurt:2002kd,Frankfurt:2003gx, Frankfurt:2003zd}. In particular, using factorization theorems at high $Q^2$, shadowing for each parton species was calculated. Calculation of gluon shadowing at low $x$ within the same framework was also presented in \cite{Tywoniuk:2007xy}. Nonetheless, the extrapolation of these results beyond the present experimental range, e.g. $x<10^{-4}$, is much less sound since unitarity can be violated for the gluons already at $x\sim 10^{-4}$ at low momentum scales \cite{Kaidalov:2003xf}.

The model \cite{Armesto:2010ee} used as basis for our present calculations is free of these shortcomings, and can safely be extrapolated down to very low $x$ in the nuclear case too. Thanks to the inclusion of scaling violations coming from QCD evolution it can also be used at high $Q^2$, a regime relevant for future hadronic colliders. While our previous work \cite{Armesto:2010ee} meticulously included extensions of the original non-perturbative formulation to the region of $x \rightarrow 1$ for both the inclusive and diffractive cross sections, we will here focus solely on the low-$x$ regime where the summation of intermediate large-mass diffractive states gives the leading contribution. Since the present analysis contains only contributions from triple-pomeron diagrams, it is therefore valid only for describing large mass diffraction -- additional ingredients would be required to properly describe low-mass diffraction and elastic scattering. Also, our model does not include anti-shadowing, see e.g. \cite{Nikolaev:1975vy,Melnitchouk:1992eu}, which is clearly seen in data for $0.1 < x < 0.3$.

The scope of this paper is as follows. In Sec.~\ref{sec:CFSKshadowing} we will describe the total $\vp A$ cross section as a multiple scattering series and re-sum it in the context of two classes of diagrams, the fan and eikonal ones. Results for shadowing at low $x$ and relatively high $Q^2$ are presented and compared with experimental data. The comparison to data at smaller $Q^2$ can be found in \cite{Armesto:2003fi}. We also calculate the nuclear modifications to the longitudinal structure function. In Sec.~\ref{sec:CFSKdiffractive} we calculate the cross section of large mass diffraction off nuclei by summing up all fan diagrams. We note that Regge factorization, which is seen to hold for diffractive DIS off protons, is violated in the nuclear case. We also calculate the ratio of diffractive and total $\vp A$ cross sections as a function of the mass of the system. Finally, we present our conclusions in Sec.~\ref{sec:Conclusions}.

\section{Nuclear shadowing of $F_2$ and $F_L$ \label{sec:CFSKshadowing}}

We proceed by calculating the total $\vp A$ cross section and, thereby, the nuclear structure function, $F_2^A$, assuming that the nucleus is made of independent nucleons. In the coherent regime, despite the complex space-time picture of the interaction, the $\vp A$ cross section can be expanded in a multiple scattering series, such as
\beq
\label{eq:MultipleScattering}
\sigma_{\vp A} = \sigma_{\vp A}^{(1)} + \sigma_{\vp A}^{(2)} + ...
\eeq
Whereas the first rescattering cross section is simply $A \sigma_{\vp N}$, the second and higher-order rescattering cross sections involve inelastic intermediate partonic states, and can be found using the AGK cutting rules \cite{Abramovsky:1973fm}. Assuming purely imaginary amplitudes, the second rescattering is simply proportional to the inclusive diffractive cross section, given by
\beq
\sigma_{\vp A}^{(2)} = -4 \pi A (A-1) \, \int d^2b \, T_A^2(b) \int^{M^2_{max}}_{M^2_{min}} \!\!\! d M^2 \left. \frac{d \sigma^\D_{\vp N}}{dM^2 dt} \right|_{t=0} \, F_A^2 (t_{min}) \;,
\eeq
where the nuclear form factor is
\be
F_A(t_{min}) = \int d^2b \, J_0 \left( b \sqrt{- t_{min}} \right) T_A(b) \; ,
\ee
and the minimal momentum transfer to the nucleus is $t_{min} \approx - m_N^2 \xP^2$. The nuclear density, $\rho_A(b,z)$, is taken to be a 2-parameter Woods-Saxon normalized to unity, with $T_A(b) = \int dz \rho_A(b,z)$.
The $t$-integrated diffractive cross section is obtained from the total reduced diffractive cross section assuming an exponential $t$-behaviour with slope $B_0$, and is given by
\beq
\label{eq:tIntegrated}
\left. \frac{d \sigma^\D_{\vp N}}{dM^2 dt} \right|_{t=0} = \frac{4 \pi^2 \alpha_{em} \, \left(B_0 - \alpha'_\Pom \ln 1\big/ \xP\right)}{Q^2 (Q^2 + M^2)} \, \xP F_{2 \D}^{(3)} \left(Q^2,\xP,\beta \right) \;,
\eeq
where the exponential slope $B_0$ is taken from \cite{Aktas:2006hx} to be $5.5$ GeV$^{-2}$. Finally, the double rescattering contribution for the deuteron takes the form
\beq
\label{eq:SecondRescattering}
\sigma_D^{(2)} = -2 \int^{t_{min}}_{-\infty} dt \, \int^{M^2_{max}}_{M^2_{min}} d M^2 \, \left. \frac{d \sigma^\D_{\vp N}}{d M^2 d t} \right|_{t=0} \, F_D (t) \,
\eeq
where $F_D(t) = e^{a t}$, $a = 40$ GeV$^{-2}$. The limits on the integral of masses in eq.~(\ref{eq:SecondRescattering}) are $M^2_{min} = 4 m_\pi^2$ and $M^2_{max}$ is found from the condition that $x_\Pom^{max} = 0.1$,\footnote{Where $x_\Pom = \left( M^2 + Q^2\right)\big/ \left( W^2 + Q^2 \right)$.} which guarantees the appearance of a measurable rapidity gap in the event.

\begin{figure}[t!]
\begin{center}
\includegraphics[width=0.4\textwidth]{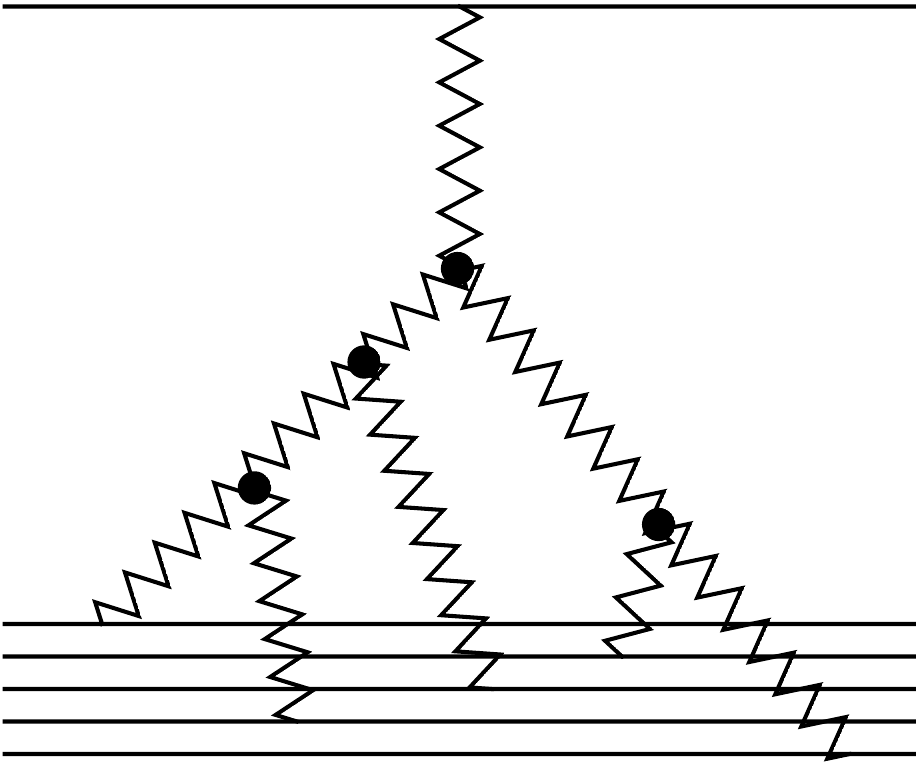}
\caption{A generic diagram in DIS off a nucleus in the Schwimmer model. Each zigzag line is considered as a full $\vp p$ amplitude.}
\label{default}
\end{center}
\end{figure}
The summation of the full rescattering series in eq.~(\ref{eq:MultipleScattering}) is model dependent.  At low $x$ we neglect effects related to isospin. Here we include two unitarization schemes. Summing all fan diagrams we arrive at the Schwimmer model \cite{Schwimmer:1975bv}, valid for collisions of a small projectile on a large target (see the Appendix for more details). Finally, the total $\vp A$ cross section is given by
\beq
\label{eq:SchwimmerModel}
\sigma_{\vp A}^{Sch} = A\sigma_{\vp p} \, \int d^2 b \, \frac{T_A(b)}{1 + (A-1) T_A(b) \, f(\xP^{max},Q^2) \big/ F_2(x,Q^2)} \;.
\eeq
On the other hand, the summation of multiple rescatterings in the eikonal unitarization gives
\beq
\label{eq:EikonalModel}
\sigma_{\vp A}^{eik} = A\sigma_{\vp p} \, \int d^2 b \, \frac{1 - \exp \left[ -2(A-1) T_A(b) \, g(\xP^{max},Q^2)\big/ F_2(x,Q^2) \right] }{2(A-1) \, f(\xP^{max},Q^2)\big/F_2(x,Q^2)} \;.
\eeq
The eikonal resummation gives stronger nuclear shadowing than the Schwimmer model, which can be checked by comparing the second non-trivial rescattering term. In eqs.~(\ref{eq:SchwimmerModel}) and (\ref{eq:EikonalModel}) the function $g(\xi,Q^2)$ is defined as 
\beq
\label{eq:ShadowingFunction}
f(\xi,Q^2) \,=\, 4 \pi \, \int^{\xi}_{x} \frac{ d \xP}{\xP} \; B(\xP) \, \xP \FD (\xP, \beta, Q^2) \, 
F_A^2(t_{min}) \;,
\eeq
where $B(\xP) = B_0 - \alpha'_\Pom \ln 1\big/ \xP$.
We define the shadowing ratio for the nuclear total cross section and, thus, the nuclear structure function, for two nuclei $A$ and $B$, as $R_{F_2} (A /B) = B \sigma_{\vp A} \big/ A \sigma_{\vp p} = B F_2^A(x,Q^2) \big/ A F^B_2(x,Q^2)$.\footnote{We automatically define $R_{F_2} (A) = F_2^A (x,Q^2) \big/ A F_2(x,Q^2)$, where $F_2$ is the structure function of a proton.}

\begin{figure}[tbp]
\begin{center}
\includegraphics[width=0.3\textwidth]{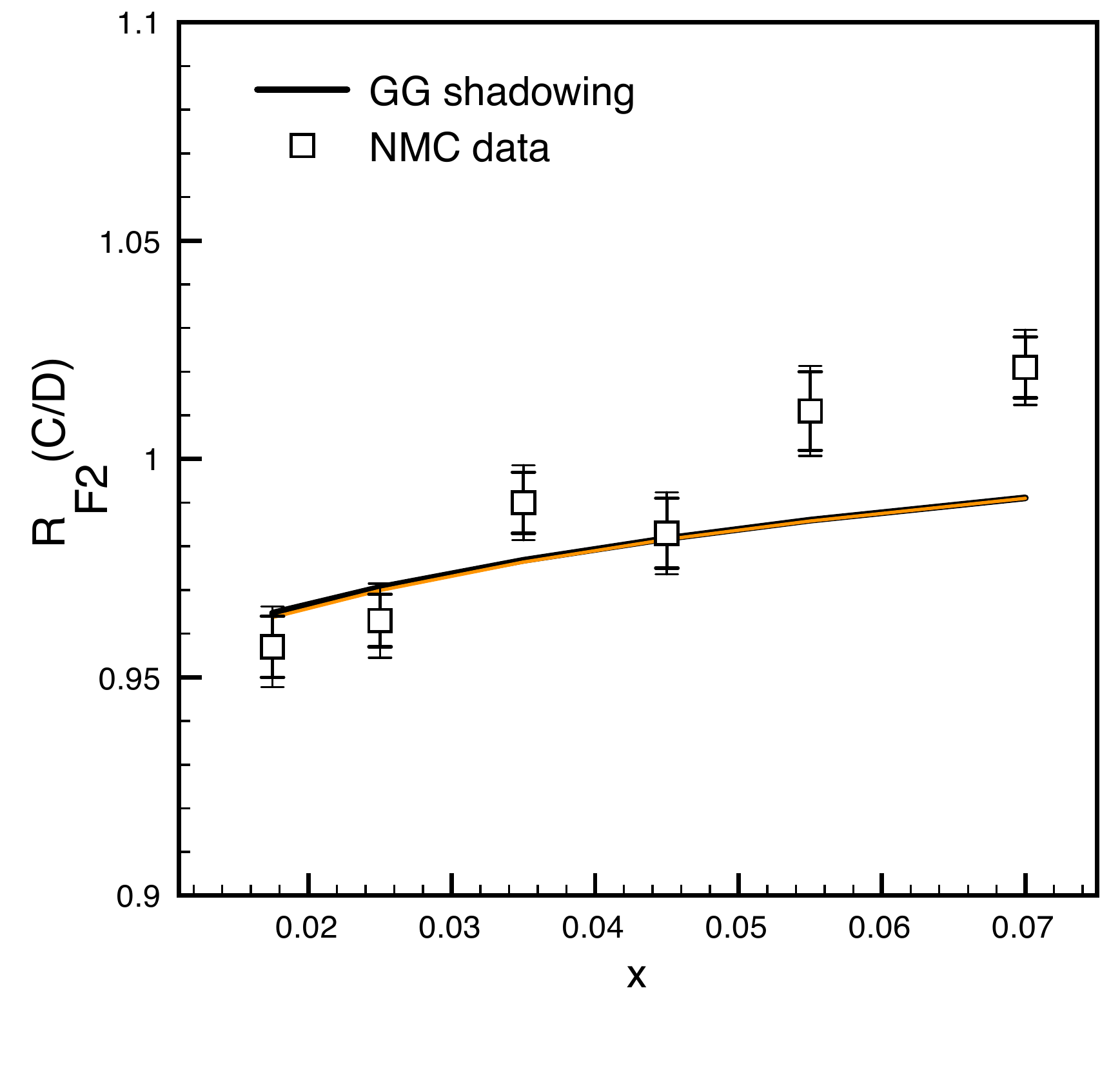}%
\includegraphics[width=0.3\textwidth]{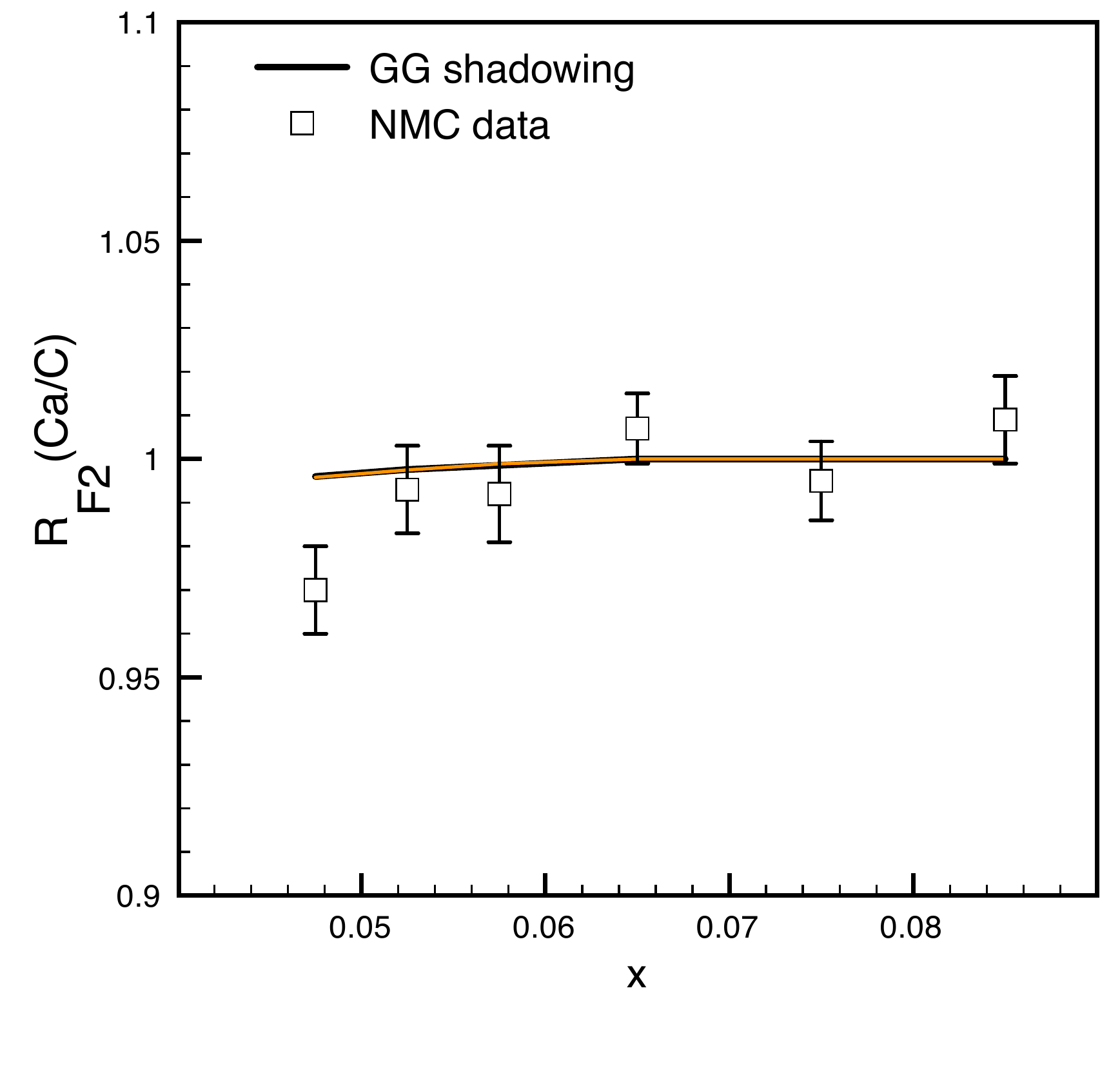}%
\includegraphics[width=0.3\textwidth]{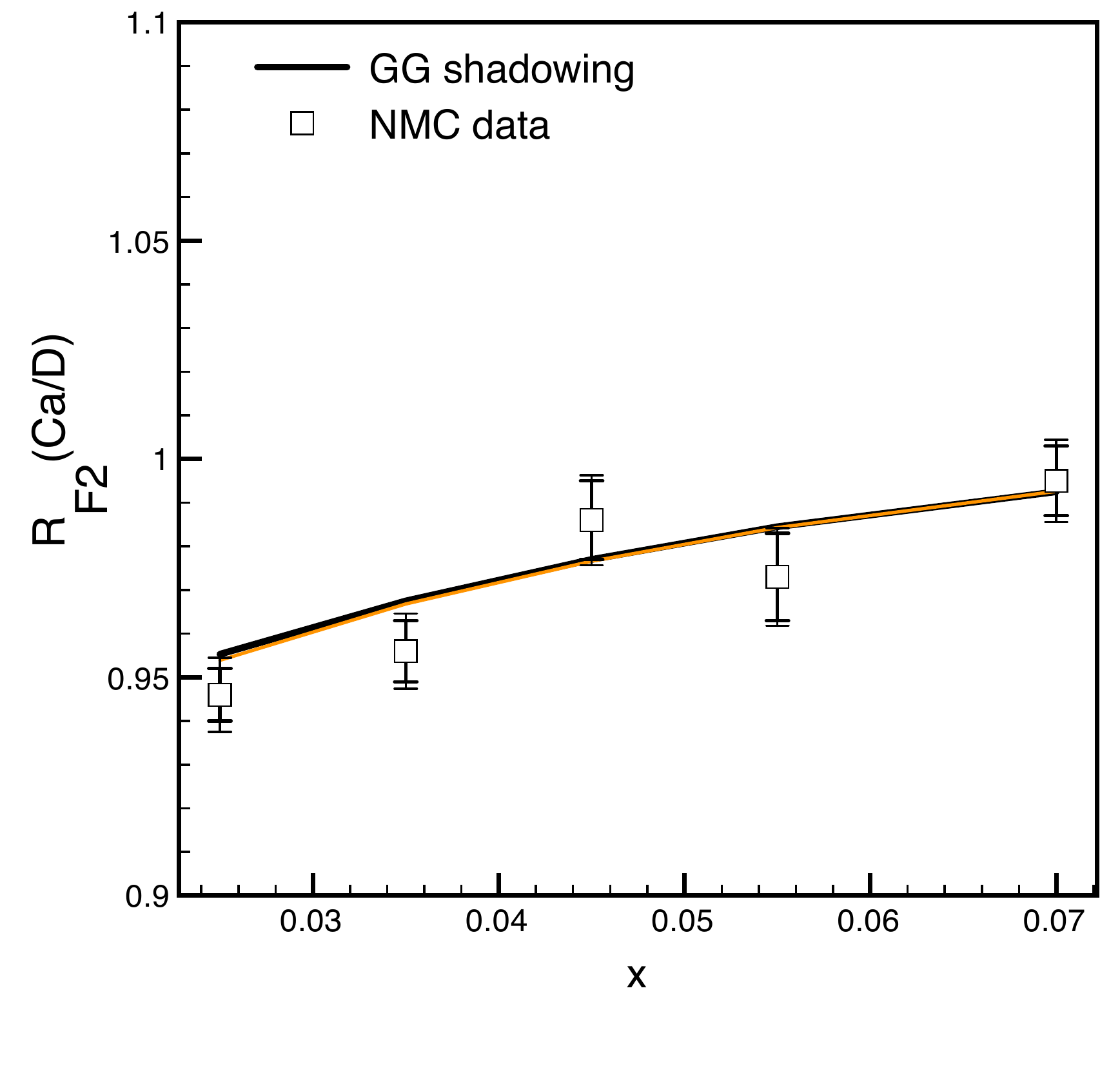}%
\caption{Nuclear shadowing compared to experimental data from \cite{Amaudruz:1995tq}. Solid (black) curves are calculated in the Schwimmer model, dashed (orange) ones in the eikonal model (note that the lines overlap).}
\label{fig:ShadExperiment1}
\end{center}
\end{figure}
Unfortunately, there is little available data for the kinematics where our model is valid, i.e. $x \leq 10^{-2}$ and $Q^2 \geq 2 $ GeV$^2$. Utilizing the results on $\xP F_{2 \D}^{(3)}(\xP,\beta,Q^2)$ and $F_2(x,Q^2)$ found in \cite{Armesto:2010ee}, we compare shadowing calculated in the Glauber-Gribov theory (the curves for the Schwimmer and eikonal unitarizations overlap) to experimental lepton-nucleus data \cite{Amaudruz:1995tq} in Fig.~\ref{fig:ShadExperiment1}. Although the model calculations are connected by solid lines, we have made calculations at the corresponding $\langle Q^2 \rangle$ of each data point.
\begin{figure}[tbp]
\begin{center}
\includegraphics[width=0.4\textwidth]{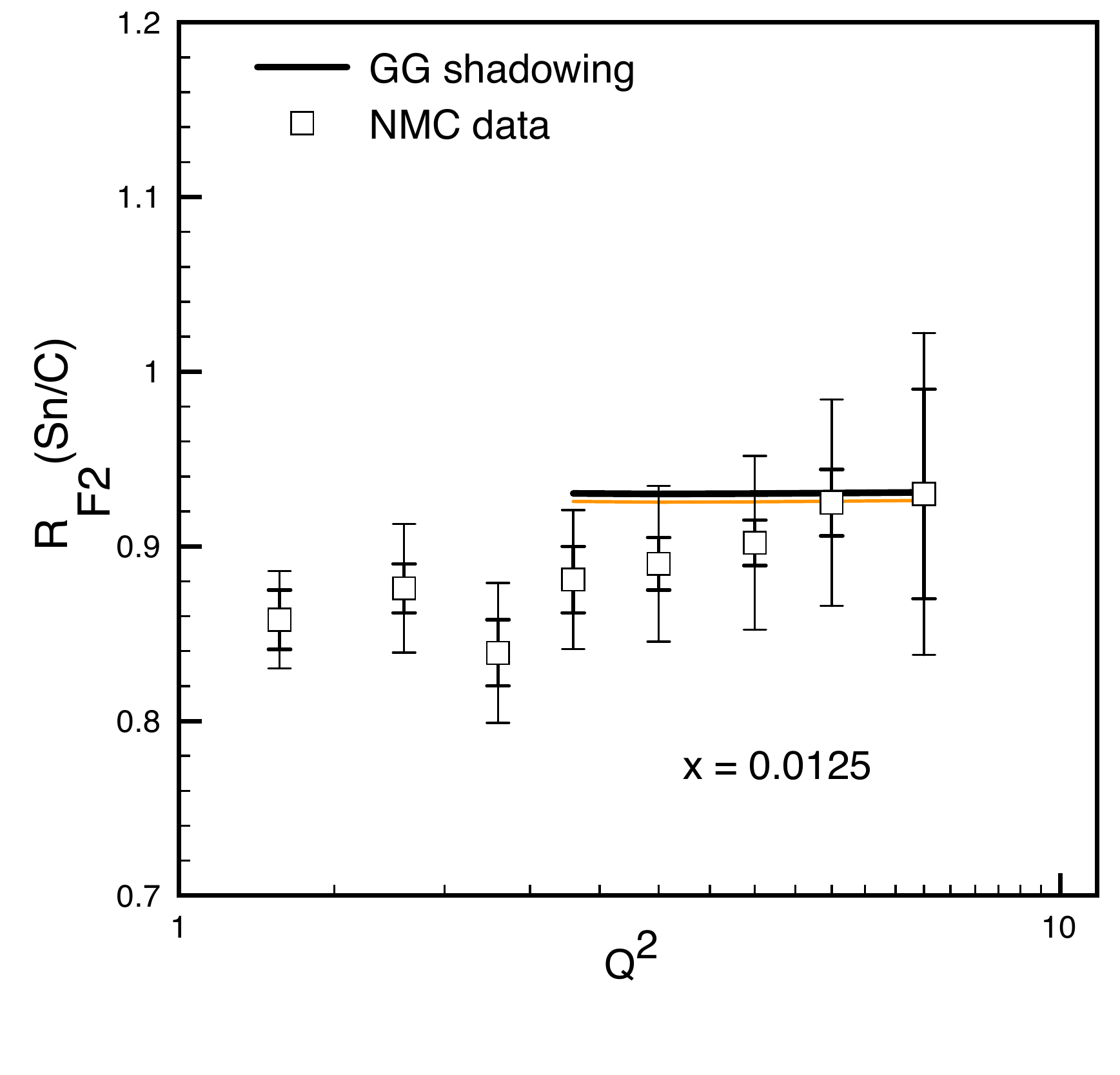}%
\includegraphics[width=0.4\textwidth]{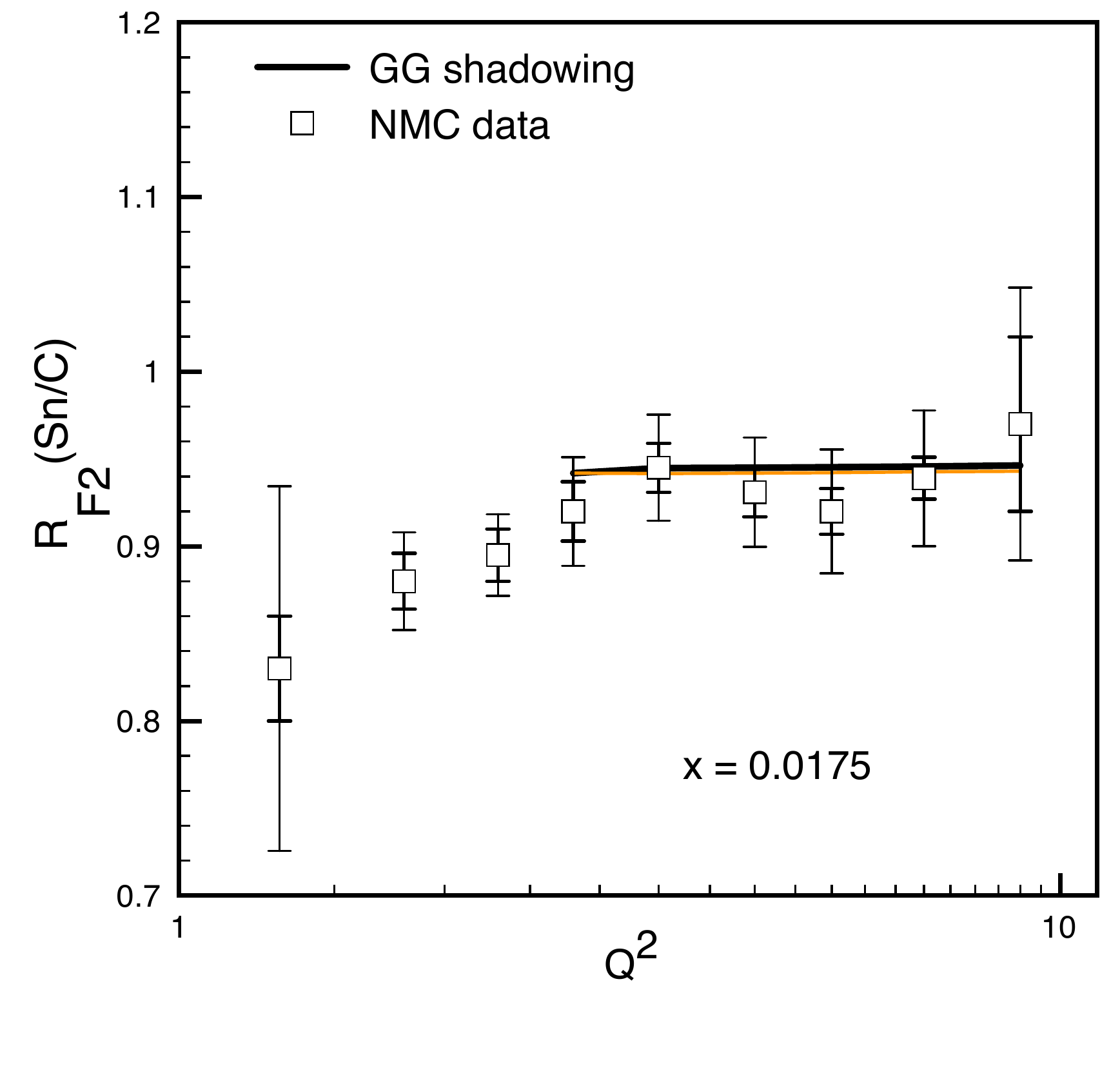}%
\caption{Nuclear shadowing compared to experimental data from \cite{Arneodo:1996ru}. Solid (black) curves are calculated in the Schwimmer model, dashed (orange) ones in the eikonal model (note that the lines overlap).}
\label{fig:ShadExperiment2}
\end{center}
\end{figure}
Furthermore, we check the $Q^2$-dependence of shadowing in Fig.~\ref{fig:ShadExperiment2} compared to experimental data from \cite{Arneodo:1996ru}.
\begin{figure}[tbp]
\begin{center}
\includegraphics[width=0.48\textwidth]{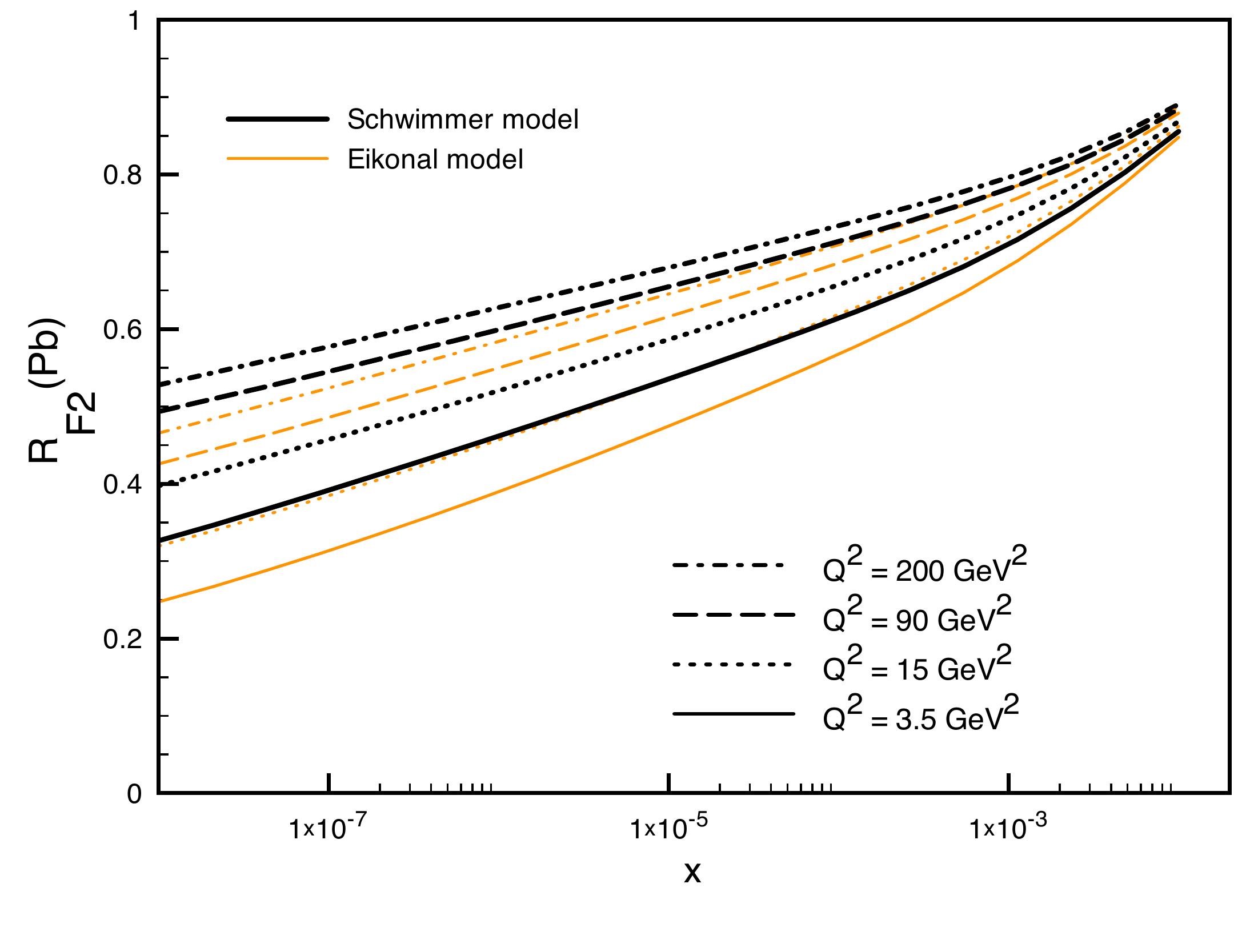}%
\includegraphics[width=0.48\textwidth]{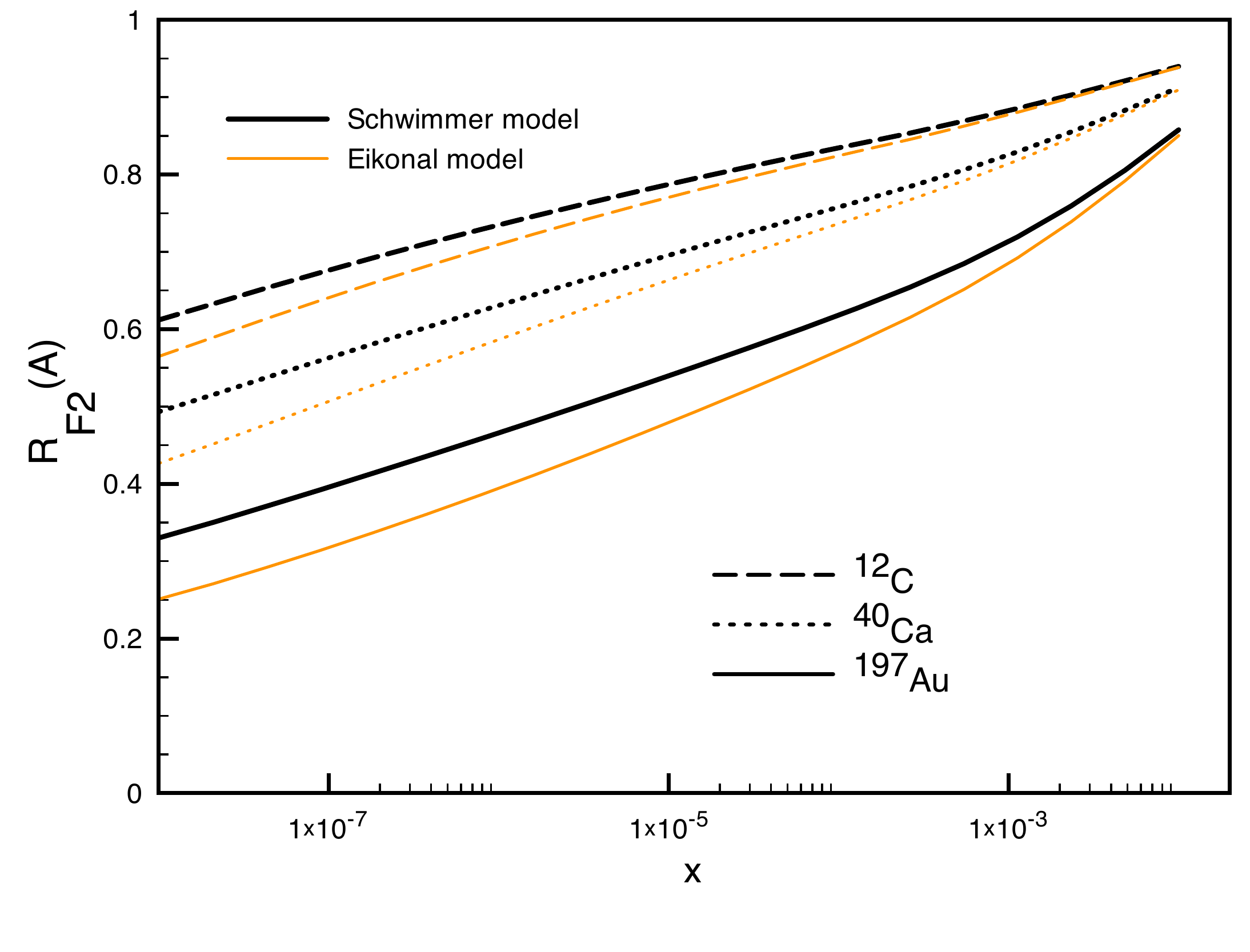}
\caption{$Q^2$-dependence of nuclear shadowing for lead (Pb) (left) and shadowing for several lighter nuclei at $Q^2 = 3.5$ GeV$^2$ (right). Upper (black) curves are calculated in the Schwimmer model, lower (orange) ones are calculated in the eikonal model.}
\label{fig:ShadPb}
\end{center}
\end{figure}
Shadowing for a lead (Pb) nucleus at several momentum scales is shown in Fig.~\ref{fig:ShadPb} (left), and the $A$-dependence is presented the right part of Fig.~\ref{fig:ShadPb}.

\begin{figure}[tbp]
\begin{center}
\includegraphics[width=0.48\textwidth]{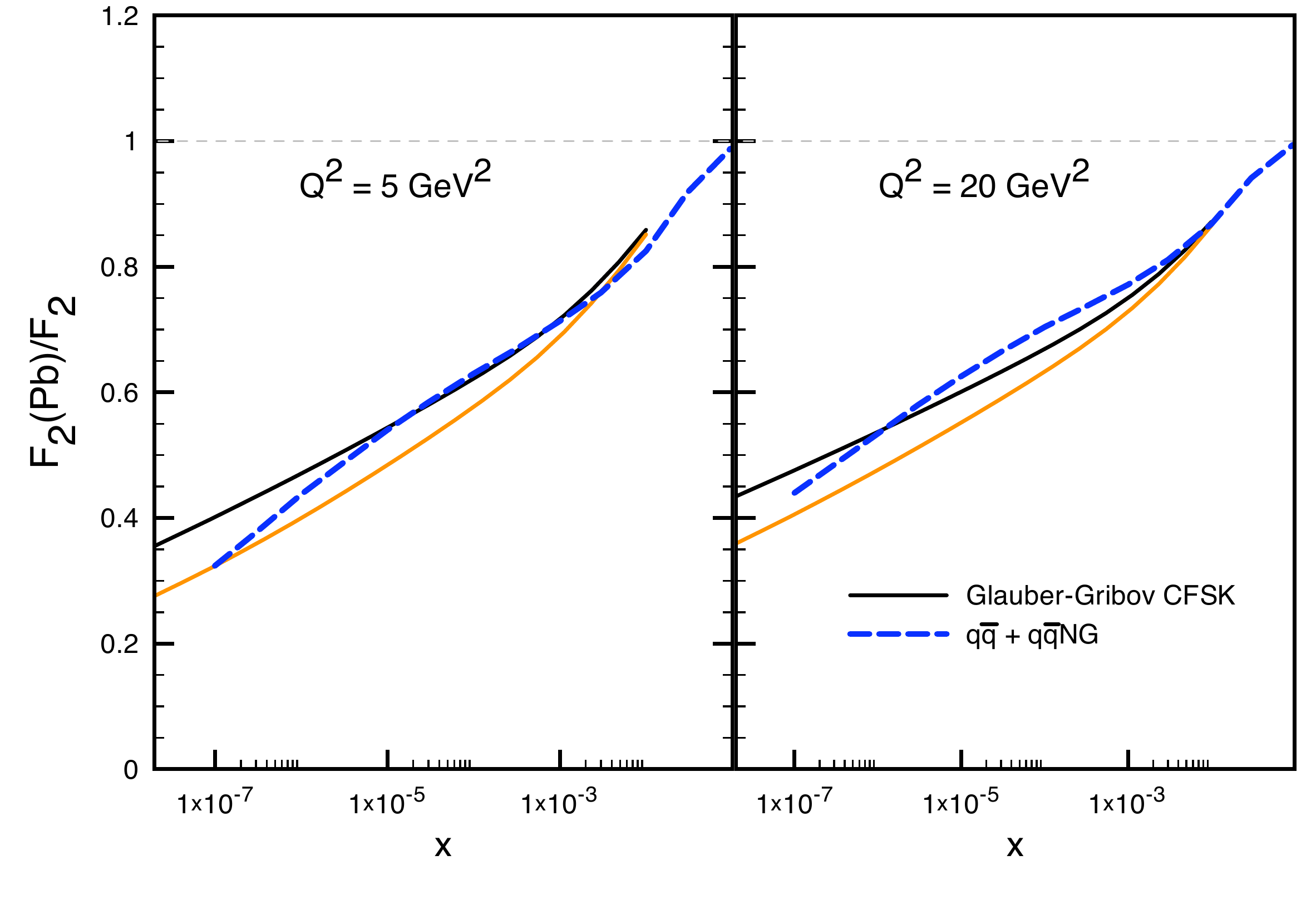}
\includegraphics[width=0.48\textwidth]{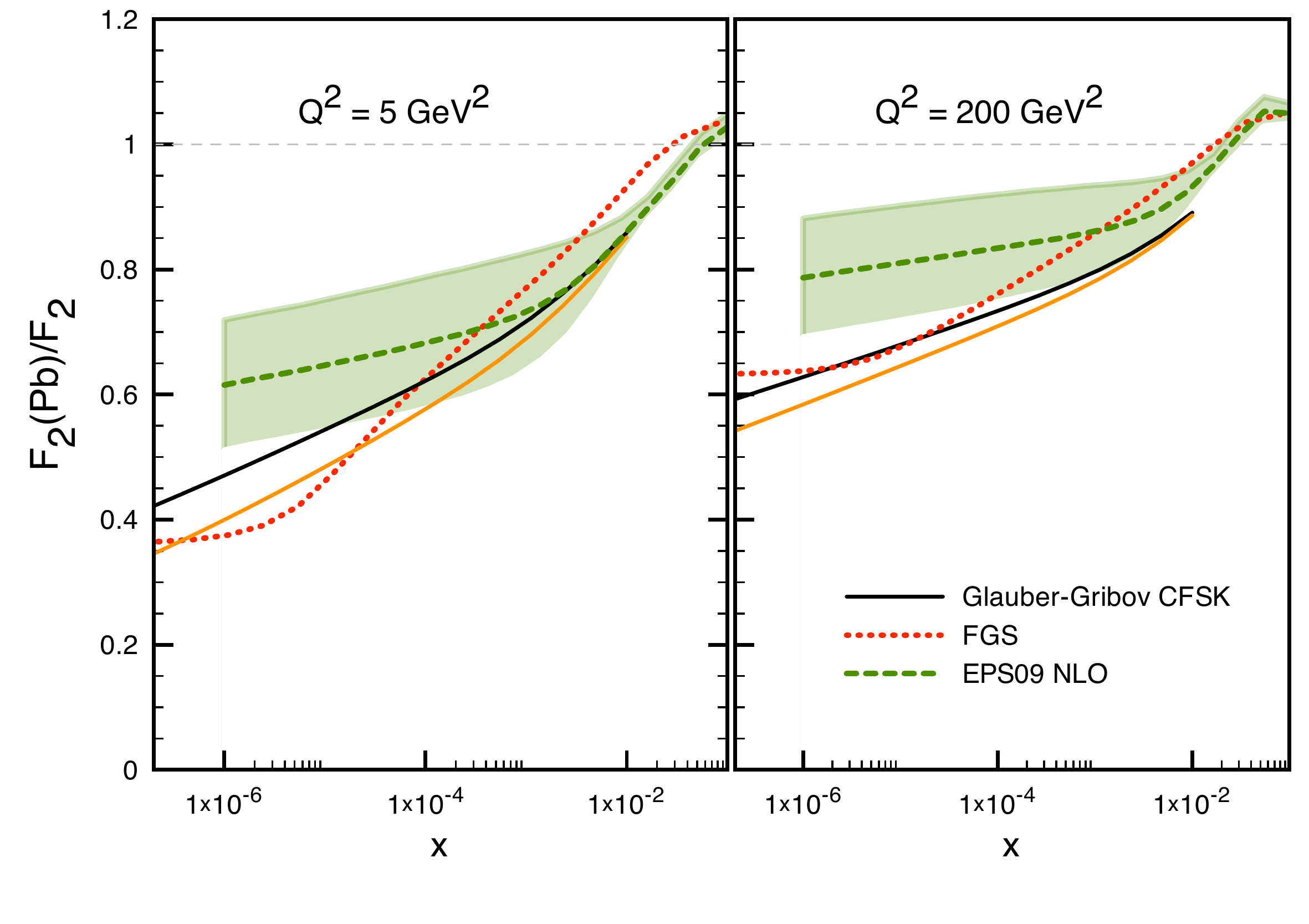}
\caption{Results for $F_2(\mbox{Pb})\big/ F_2$ compared to the dipole calculation in \cite{Kopeliovich:2008ek} (left), the EPS09 NLO parameterization \cite{Eskola:2009uj} (right) and the FGS model \cite{Frankfurt:2003zd} (both). Solid upper (black) curves are calculated in the Schwimmer model, lower (orange) ones are calculated in the eikonal model.}
\label{fig:ShadModels}
\end{center}
\end{figure}
A comparison to several models of nuclear modifications is presented in Fig.~\ref{fig:ShadModels}. Our model is in good agreement with the recent calculation within the dipole model calculation of \cite{Kopeliovich:2008ek}, where rescatterings of the full $\qqbar$+N$g$ fluctuation is taken into account, see left plots in Fig.~\ref{fig:ShadModels}. The higher Fock-states of the dipole correspond to the summation of triple-pomeron diagrams in our approach.

Furthermore, we obtain reasonable consistency at low $x$ between our result and the calculations presented in \cite{Frankfurt:2003zd}, right part of Fig.~\ref{fig:ShadModels}. Different from our work, in \cite{Frankfurt:2003zd} the input partonic densities of the proton and pomeron were taken from perturbative fits of experimental data at NLO. Other differences stem mainly from the treatment of the large-$x$ region, e.g. the parameter ${\xP}_{max}$ is put to 0.03 in \cite{Frankfurt:2003zd} and strong anti-shadowing is included in the parameterization of the gluons stemming from the momentum sum rule of the nucleus. The latter corresponds to secondary reggeon exchanges \cite{Melnitchouk:1992eu} which are neglected in our model. The agreement at high-$Q^2$ is also reasonable. Finally, we compare our calculations to the recent parameterization of shadowing extracted from a NLO DGLAP analysis of nuclear data \cite{Eskola:2009uj} in the right part of Fig.~\ref{fig:ShadModels}. The errors of the nuclear parameterization are also indicated. We find large differences between the approaches at low momentum scales at $x < 10^{-2}$, which are solely due to the choice of initial conditions. The model in \cite{Eskola:2009uj} predicts a significantly smaller nuclear shadowing at $x < 10^{-4}$. The discrepancy persists at higher $Q^2$.

It is worth pointing out the different treatment of $Q^2$ evolution between the different models. Our treatment of the Glauber-Gribov rescattering series is valid for both soft and hard processes at low-$x$ and involves effectively all twist contributions. Nevertheless, it was demonstrated for the non-perturbative regime in \cite{Armesto:2003fi} that the leading-twist contribution is by far the strongest. Note the difference with \cite{Frankfurt:2003zd}, where the nuclear PDFs at some initial scale were used as the input to DGLAP evolution. The same approach was applied in \cite{Eskola:2009uj}, where initial nuclear PDFs were fitted. Finally, the dipole model calculation \cite{Kopeliovich:2008ek} also account for some scaling violations. The $Q^2$ evolution of the nuclear structure function is clearly logarithmic in all the discussed models. 

\begin{figure}[tbp]
\begin{center}
\includegraphics[width=0.55\textwidth]{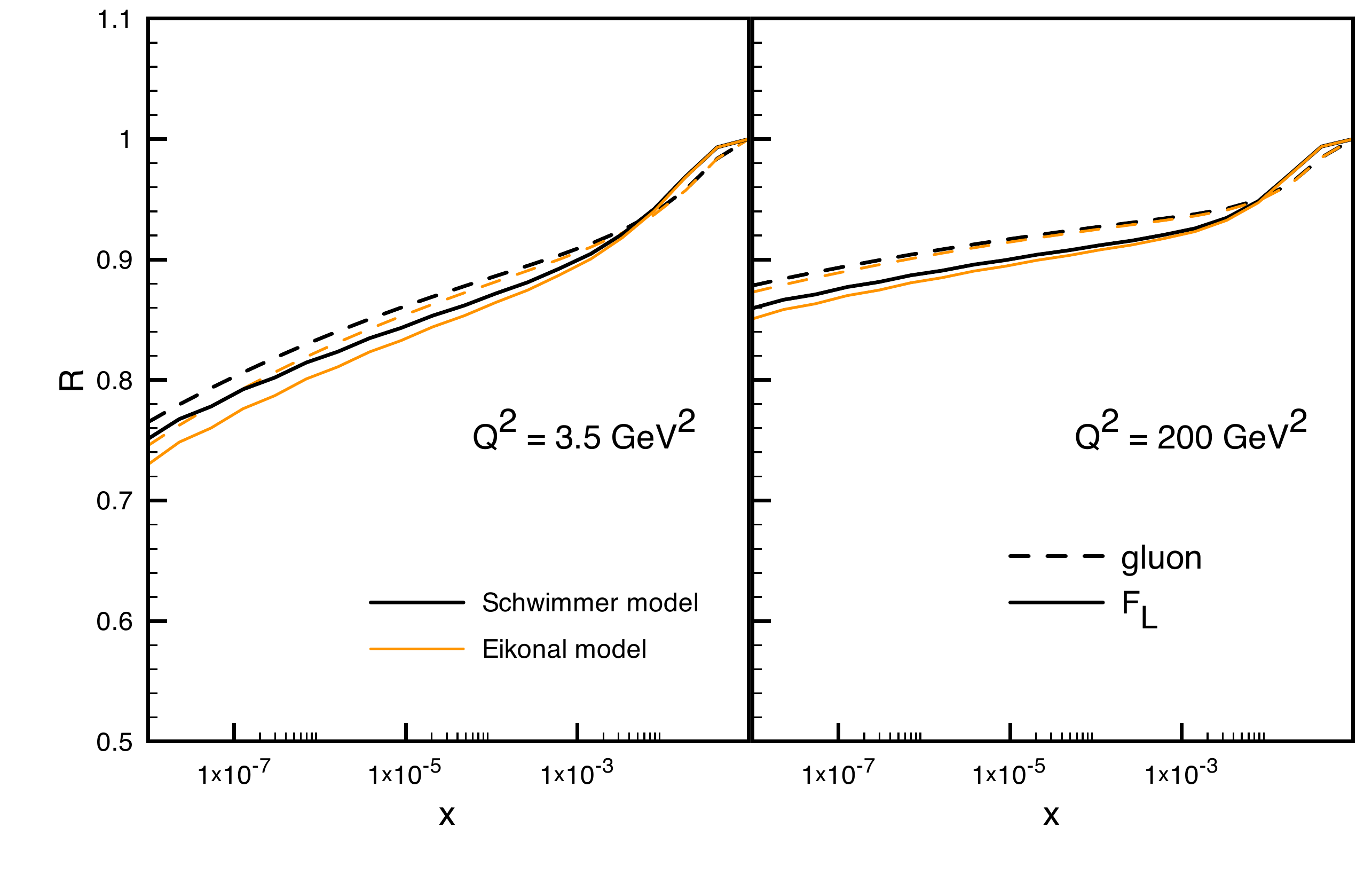}
\caption{Shadowing of the nuclear $F_2$ (dotted) and $F_L$ (solid) structure functions together with the gluon nuclear distribution (dashed curves). Upper (black) curves are calculated in the Schwimmer model, lower (orange) ones are calculated in the eikonal model.}
\label{fig:ShadFL}
\end{center}
\end{figure}
We continue our study by analyzing the longitudinal part of the total cross section. In the naive quark-parton model, the Callan-Gross relation gives that $F_L = 0$. This is not the case when we include radiation of quarks and gluons. The longitudinal structure function can readily be calculated using the Altarelli-Martinelli formula \cite{Altarelli:1978tq}
\beq
\label{eq:AltarelliMartinelli}
F_L (x,Q^2) = \frac{2 \alpha_S(Q^2)}{\pi} \int^1_x \frac{d y}{y} \, \left( \frac{x}{y} \right)^2 \left( \sum^{N_f}_{i = 1} e^2_i \left( 1- \frac{x}{y} \right) y g(y,Q^2) \,+\, \frac{2}{3} F_2 (y,Q^2) \right) \;,
\eeq
relating $F_L$ quantities at LO on the right-hand side of the equation, where $N_f$ is the number of active quark flavours. Due to large magnitude of the gluon PDF compared to quarks in the proton, this quantity is dominated by the gluon component. The corresponding relation for the nuclear case is analogously found, so that we can define the nuclear suppression factor, $R_{F_L}= F_L^A(x,Q^2) \big/ A F_L(x,Q^2)$, of the longitudinal structure function as
\beq
R_{F_L} = \frac{\int^1_x \frac{d y}{y} \, \left( \frac{x}{y} \right)^2 \left( \sum^{N_f}_{i = 1} e^2_i \left( 1- \frac{x}{y} \right) y \, R_g(y,Q^2)g(y,Q^2) + \frac{2}{3} R_{F_2}(y,Q^2)F_2 (y,Q^2) \right)}{\int^1_x \frac{d y}{y} \, \left( \frac{x}{y} \right)^2 \left( \sum^{N_f}_{i = 1} e^2_i \left( 1- \frac{x}{y} \right) y g(y,Q^2) \,+\, \frac{2}{3} F_2 (y,Q^2) \right)} \;,
\eeq
where we calculate with $N_f=4$. In the equation above, $R_g$ denotes gluon shadowing and is calculated analogously to $R_{F_2}$ by substituting $f(\xi,Q^2)$ by
\beq
f (\xi,Q^2) \,\rightarrow\, 4 \pi \, \int^{\xi}_{x} \frac{ d \xP}{\xP} \; B(\xP) \, \xP \overline{f}_{\Pom/p}(\xP) \, \beta g_\Pom(\beta,Q^2) \, 
F_A^2(t_{min}) \;,
\eeq
where $\overline{f}_{\Pom/p}(\xP)$ is the $t$-integrated pomeron flux factor and $g_{\Pom}(\beta,Q^2)$ is the pomeron gluon distribution, and $F_2(x,Q^2) \rightarrow x g(x,Q^2)$ in eq.~(\ref{eq:SchwimmerModel}). Both of these quantities are taken from \cite{Armesto:2010ee}.

Results for $R_{F_L}(x,Q^2)$ together with $R_g(x,Q^2)$ are shown in Fig.~\ref{fig:ShadFL}. Due to the small gluon shadowing, nuclear shadowing of the longitudinal structure function is almost the same as for the total cross section. The smallness of $R_g$, on the other hand, is caused by the rapid small-$x$ growth of the gluon PDF in the proton at LO \cite{Armesto:2010ee}. Note that studies at NLO \cite{Frankfurt:2002kd, Frankfurt:2003gx,Tywoniuk:2007xy} infer a more pronounced gluon shadowing, resulting in a significantly smaller $R_{F_L}$ than shown in Fig.~\ref{fig:ShadFL}, see e.g. \cite{Frankfurt:2003gx}.

\section{Diffraction off nuclei in the Schwimmer model \label{sec:CFSKdiffractive}}

Diffraction in DIS provides valuable complementary information to the total cross section as it probes the parton distributions at large impact parameters and involve a large soft component. Nuclear DIS possess a rich structure and have recently triggered significant activity \cite{Frankfurt:1993qi,Nikolaev:1995xu,Gotsman:1999vt,Levin:2002fj,Kaidalov:2003vg,Frankfurt:2003gx,Kugeratski:2005ck,Nikolaev:2006mh,Kowalski:2007rw,Kowalski:2008sa}. Generally speaking, one naturally expects a large diffractive cross section in the nuclear case due to the large probability of rescattering. Several groups predict a strong enhancement of the fraction of diffractive events, up to 30-40\% \cite{Nikolaev:1995xu,Kugeratski:2005ck,Kowalski:2008sa} (compared to 15\% in $\vp p$ at HERA), in the nuclear case. Note that most of this enhancement derive from the elastic component, while scattering of higher-order Fock-states are found to be suppressed \cite{Frankfurt:2003gx,Nikolaev:2006mh, Kowalski:2008sa}. This is in line with the standard Good-Walker picture of diffraction, and stems from the fact that as the target gets closer to the black disc limit for central and semi-central impact parameters at high energies, only the periphery of the nucleus contributes to inelastic diffraction.

\begin{figure}[t!]
\begin{center}
\includegraphics[width=0.6\textwidth]{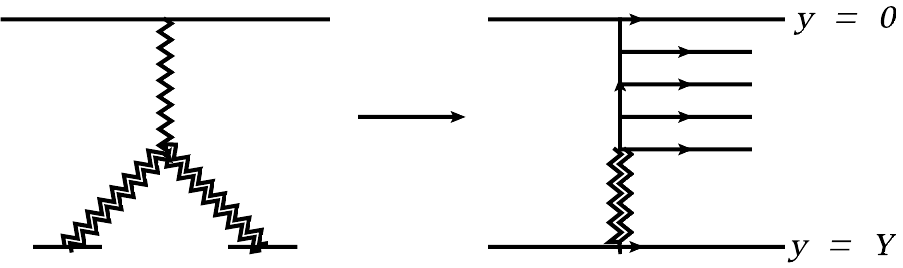}
\caption{The single diffractive cross section in the Schwimmer model. Note that the double zig-zag lines at the bottom sums all fan diagram contributions. The top incoming line corresponds to a virtual photon, while the lower one to a target nucleon/nucleus.}
\label{fig:SchwimmerDiffraction}
\end{center}
\end{figure}
In this work, we focus uniquely on the diffractive processes where a large mass is being produced. In this limit, the summation of fan diagrams will dominate the cross section. Although the mass spectrum is steeply falling, $\sim 1\big/ M^2$, the experimental reach of planned lepton-hadron colliders, such as e.g. eRHIC \cite{Deshpande:2005wd} and LHeC \cite{lhec2}, will be sensitive to these effects. Nuclear diffraction at $\beta \to 1$ is large, and an extension of the model to this region, where rescatterings are very important, would be e.g. to include vector-meson dominance, in the spirit of \cite{Bauer:1977iq,Piller:1995kh}.  Besides, we do not include the possibility of intermediate excitations of target protons, e.g. by introducing a two-channel model as in \cite{Kaidalov:2003vg}. Additionally, we consider totally coherent diffraction, i.e. assuming that the nucleus stays completely intact after the interaction with the projectile. In this case, the momentum transfer to the nucleus must be very small, $|t| \sim 1 \big/ R_A^2$. This situation seems more easily tractable from the theoretical point of view, whereas the calculation of incoherent diffraction involves calculation of survival factors with process dependent absorptive cross sections \cite{Kaidalov:2003vg} which are {\it a priori} unknown. One the other hand, the full $t$-dependence of diffraction off nuclei would have to be modeled including break-up mechanisms and multi-fragmentation. Measuring the intact nucleus at small $t$ at electron-ion colliders will be challenging \cite{Deshpande:2005wd}. Efforts to extend the present analysis of nuclear diffraction along these lines are under way.

After resumming all fan diagram exchanges, see Fig.~\ref{fig:SchwimmerDiffraction} (the derivation can be found in the Appendix), the single-diffractive $\vp A$ cross section at a given center-of-mass energy, $W^2$ (where $x \simeq Q^2 \big/ W^2 + Q^2$), and scale, $Q^2$, as a function of the diffractively produced mass, $M^2$, is given by \cite{Bondarenko:2000uv, Boreskov:2005ee}
\beq
\label{eq:CoherentDiffraction}
\frac{d \sigma^\D_{\vp A}}{d M^2} = \frac{d \sigma^\D_{\vp N}}{d M^2} \int d^2 b \frac{4\pi  B_0  \, A^2 T_A^2(b) 
}{ \left[ 1 + A T_A(b) \left( 2 f(\xP^{max},Q^2) - f(x/\beta,Q^2) \right)\Big/  F_2(x,Q^2) \right]^2} \;,
\eeq
where $f(\xi,Q^2)$ is defined in eq.~(\ref{eq:ShadowingFunction}) and we have assumed an exponential $t$-slope of the diffractive $\vp p$ cross section with slope parameter $B_0$. Clearly, eq.~(\ref{eq:CoherentDiffraction}) does not factorize the dependence of kinematical variables related to the emission of a (generalized) pomeron from the target (typically $\xP$ and $t$) from those related to its subsequent scattering with the $\vp$-projectile (typically $\beta = Q^2\big/ M^2 + Q^2$ and $Q^2$). This indicates that Regge factorization \cite{Ingelman:1984ns}, which is roughly valid in a large kinematical range in diffractive DIS off protons \cite{Armesto:2010ee}, is more strikingly violated for nuclei due to the stronger rescattering in the latter case.

In fact, a peculiar feature can be deduced from eq.~(\ref{eq:CoherentDiffraction}), or simply by looking at Fig.~\ref{fig:SchwimmerDiffraction}, namely that shadowing effects tend to vanish with increasing of the diffractively produced mass. This is because, as long as we demand that the nucleus stays intact, the phase space available for rescattering is reduced with increasing $M^2$. 

\begin{figure}[t!]
\begin{center}
\includegraphics[width=0.55\textwidth]{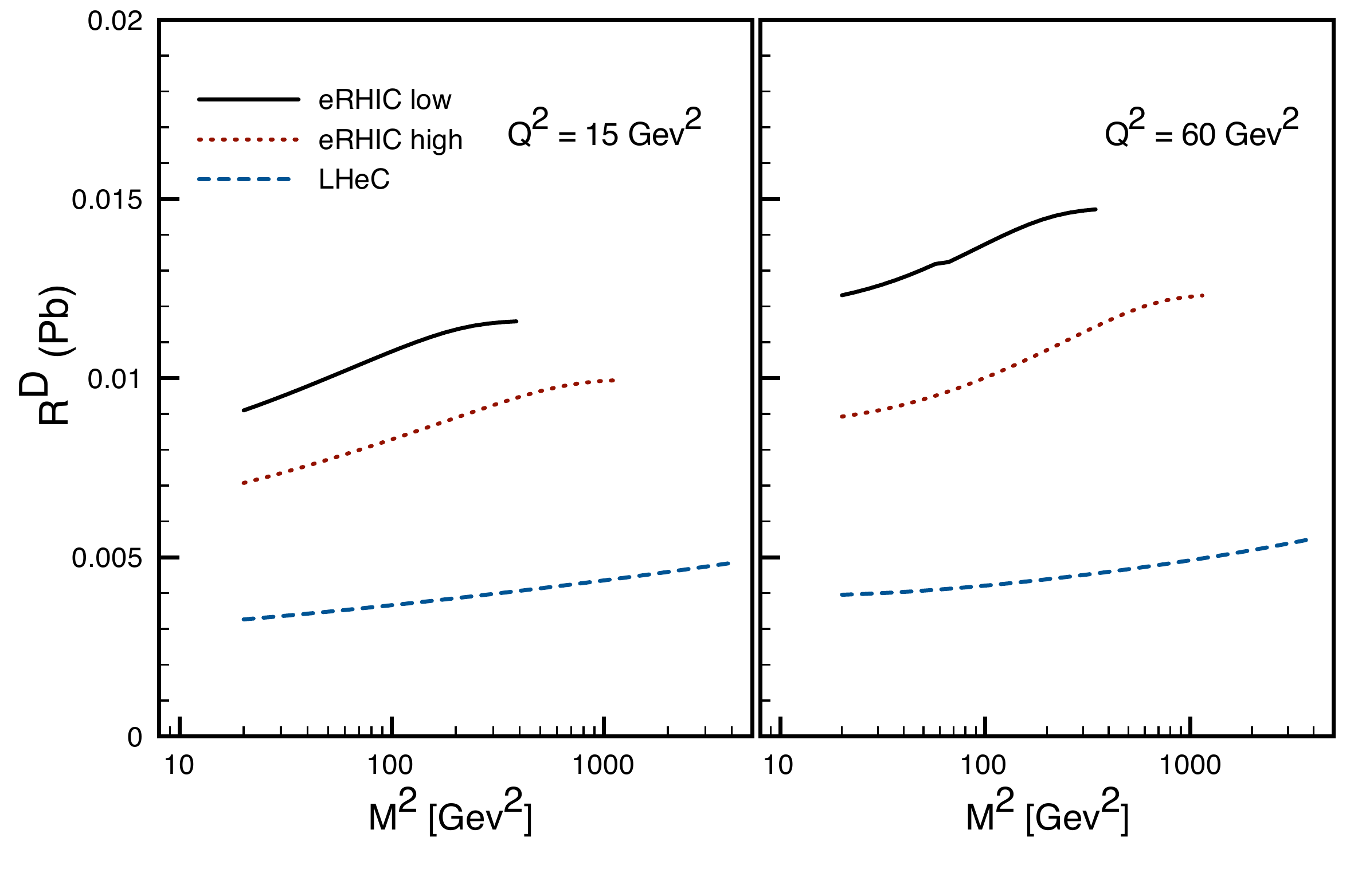}%
\caption{Nuclear diffractive ratio in the Schwimmer model with CFSK PDFs.}
\label{fig:RdiffvsMass}
\end{center}
\end{figure}
\begin{figure}[t!]
\begin{center}
\includegraphics[width=0.45\textwidth]{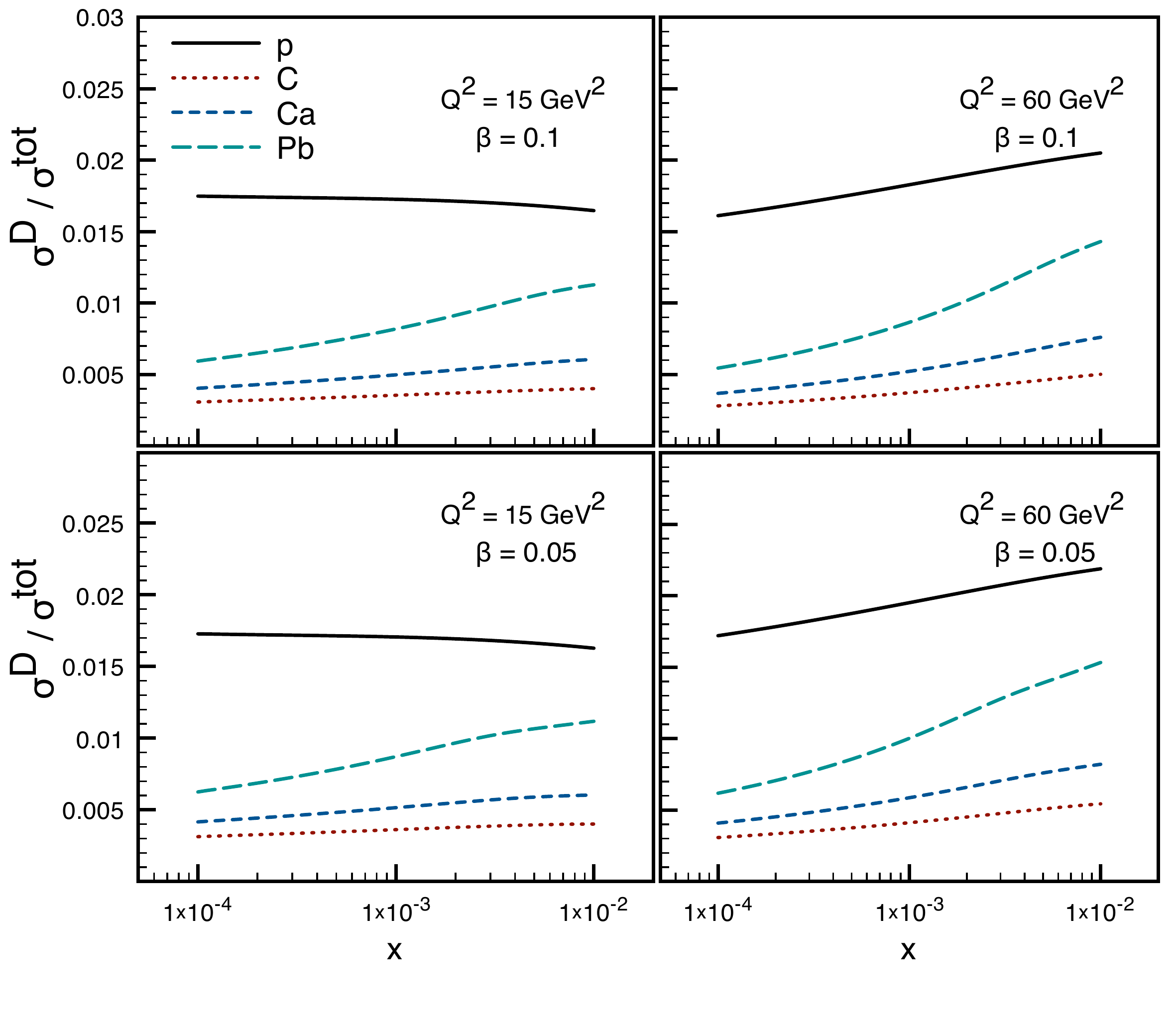}
\caption{Nuclear diffractive-to-inclusive ratio in the Schwimmer model with CFSK PDFs.}
\label{fig:SigmaDoverSigmaTot}
\end{center}
\end{figure}
We define the nuclear diffractive ratio in the standard form as
\beq
R_\D(s,M^2,Q^2) = \frac{d \sigma_{\vp A}^\D}{d M^2} \Bigg/ A^2 \, \frac{d \sigma_{\vp N}^\D}{d M^2} \;,
\eeq
for coherent diffraction.\footnote{Note that in the absence of nuclear effects (and assuming a Gaussian density profile for the nucleus) $R_\D = 2 B_0 \big/ R_A^2 \simeq A^{-2/3}$ which is the case when $R_p^2 = 2 B_0$. Hence, actually one should compare the quantity $A^{2/3} R_{\D}$ to unity.} We plot the suppression factor for Pb as a function of the diffractively produced mass for two different momentum transfers in Fig.~\ref{fig:RdiffvsMass} for three different scenarios: (i) "eRHIC low" -- assuming a beam energy of $E_A = 100 $ GeV/$c$ per nucleon for the nucleus side and an electron beam energy of $E_e = 10$ GeV/$c$; (ii) "eRHIC high" -- the same nucleus beam energy as in the previous case and $E_e = 30$ GeV/$c$; and (iii) "LHeC" -- with $E_A = 2750$ GeV/$c$ per nucleon and $E_e = 100$ GeV/$c$. We stress that this is the suppression solely due to coherent multiple rescattering \cite{Frankfurt:2003gx, Nikolaev:2006mh}.

In Fig.~\ref{fig:SigmaDoverSigmaTot} we plot the ratio $\sigma^\D \big/ \sigma^{tot}$ for protons and three nuclei ($^{\mbox{\scriptsize12}}$C, $^{\mbox{\scriptsize40}}$Ca and $^{\mbox{\scriptsize208}}$Pb) for a fixed value of $\beta$ and momentum transfer $Q^2$ as a function of Bjorken $x$. For such values of $\beta$, where we safely can assume the dominance of fan diagrams ($M^2 > 135$ GeV$^2$), we observe the suspected suppression compared to the proton case. It is also worth noticing the different behavior with $x$ of the ratio $\sigma^\D \big/ \sigma^{tot}$ for protons and nuclei at low momentum scales $Q^2$. 

Note that we systematically find a nuclear suppression of diffraction, to be contrasted with the enhancement found for smaller masses in \cite{Nikolaev:1995xu,Kugeratski:2005ck,Kowalski:2008sa}.

\section{Conclusions \label{sec:Conclusions}}

Shadowing of the nuclear structure function was calculated in the Glauber-Gribov model down to very low-$x$ and high-$Q^2$ utilizing the results on inclusive and diffractive cross sections in $\vp p$ scattering. The agreement with data is good, and comparison with predictions from other models at low-$x$ seems to be fairly consistent, both regarding the magnitude and $Q^2$-dependence, except that nuclear shadowing of $F^A_2$ is found to be significantly stronger at low $x$ than in \cite{Eskola:2009uj}. Strong shadowing effects are expected at future high-energy electron-nucleus colliders and heavy-ion collisions at LHC \cite{Capella:1999kv} . On the other hand, weak shadowing of the gluons, owing to the large gluon density of the proton at low $x$ and $Q^2$ found in  \cite{Armesto:2010ee}, lead to significantly smaller nuclear effects on $F_L^A$, in contrast to \cite{Frankfurt:2003gx}. Calculating high-mass coherent diffraction off nuclei from fan diagram resummation, we found a large suppression solely due to coherent scattering. Also, Regge factorization is strongly broken in the nuclear case.

The program of extending the above $\vp A$ analysis to the regime of large $x$, and low-mass diffraction, poses several difficulties that are not yet unambiguously resolved in the literature. In the inclusive case, vector-meson dominance and, more seriously, reggeon exchanges with the target are not included. While the former can be straightforwardly incorporated (see e.g. \cite{Frankfurt:2003zd}), the inclusion of the latter is related to anti-shadowing (the second rescattering term is positive) and would involve a much more careful treatment of the multiple scattering series. For diffraction off nuclei, the most serious challenge is posed by the treatment of low-energy nuclear effects related to the break-up of the nucleus -- a necessary ingredient for a proper comparison to an experimental situation. These lacks can only be amended through a full-fledged Monte-Carlo procedure. An estimate of fully-incoherent diffractive scattering was presented in \cite{Kaidalov:2003vg,Kowalski:2008sa}.

\subsection*{Acknowledgements}
K.T. would like to thank K.~Boreskov,  V.~Khoze, T.~Lappi, C.~Pajares and R.~Venugopalan for interesting discussions and H.~Paukkunen for comments. The work of N.A., C.A.S. and K.T. has been supported by Ministerio de Ciencia e Innovaci\'on of Spain under project FPA2008-01177 and contracts Ram\'on y Cajal (NA and CAS), by Xunta de Galicia (Conseller\'{\i}a de Educaci\'on) and through grant PGIDIT07PXIB206\-126PR (NA and KT), grant INCITE08PXIB296116PR (CAS) and European Commission grant PERG02-GA-2007-224770, and by the Spanish Consolider-Ingenio 2010 Programme CPAN (CSD2007-00042) (NA and CAS).

\appendix

\section{Derivation of the nuclear total and diffractive cross section in the Schwimmer model \label{app:SchwimmerModel}}

The cross section for diffractive production of a system with large mass in hadronic reactions is found by analyzing the contribution from triple-pomeron diagrams. In the situation of asymmetric collisions at high energies, e.g. DIS off nucleons and nuclei, diagrams with a single pomeron emitted from the projectile with its subsequent branchings give the largest contribution, i.e. that enhanced by the product of $3\Pom$-couplings and $A^{1/3}$. The resummation of all triple-pomeron branchings, the so-called fan diagrams, was first carried out in \cite{Schwimmer:1975bv} and more recently studied in \cite{Bondarenko:2000uv, Boreskov:2005ee}, where expressions for total, inclusive and single-diffractive cross sections can be found together with probabilities for large rapidity gap survival. The fan resummation with $n \rightarrow m$ pomeron vertices was presented in \cite{Ostapchenko:2006vr}. Here we recall some key results.

First, let us define a quantity which will prove helpful in describing the fan-diagram resummation. Consider scattering off a proton by means of a single triple-pomeron diagram. Assuming some maximal allowed mass, or a minimal rapidity gap, $y_M \equiv Y-y = \ln \left( M^2 \big/ s_0 \right)$, the expression is given by
\beq
\sigma^\D_{3\Pom}(\vp N \rightarrow X N) &=& \int^{Y-y_M}_0 dy d^2b d^2b' \, g^2_{\Pom N} (b) g_{\vp \Pom} (b-b') \, G_{3\Pom} e^{\Delta y} e^{2\Delta (Y-y)} \\
&=& g_{\vp \Pom} g_{\Pom N}(b=0) e^{\Delta Y} \,\cdot\, \kappa_N(Y,y_M)  \nonumber
\eeq
where we have assumed Gaussian impact parameter profiles. The first part in the expression above is the expression for a simple one-pomeron exchange while
\beq
\kappa_N (Y,y_M) = \frac{g_{\Pom N} G_{3\Pom}}{\pi \lambda_N} \frac{1}{\Delta} \left( e^{\Delta (Y-y_M)} - 1 \right) \;,
\eeq
where $G_{3 \Pom}$ is the triple-pomeron vertex. Due to the smallness of this parameter, found in phenomenological studies, $\kappa_N$ is usually considered negligible. Yet, in the nuclear case it is enhanced by the mass number, $\kappa_A(Y,y_M,b) = A T_A(b) \kappa_N(Y,y_M)$ (instead of the usual Gaussian parameterization, we will use the Woods-Saxon profile for the nuclear case) and is therefore sizeable. On the other hand
\beq
\sigma^\D_{3\Pom}(\vp N \rightarrow X N) &=& \int_{M_{min}^2}^{M^2} \!\!\!\! d M'^2  \,\left. \frac{d \sigma^\D_{3\Pom}}{d M'^2 \, dt} \right|_{t=0} \;,
\eeq
where $M^2_{min}=4 m_\pi^2$. In the following we will therefore make the following extension
\beq
\kappa_A(Y,y_M,b) &\propto& \frac{A T_A(b)}{\sigma_{\Pom} (Y)} \,  \int_{M_{min}^2}^{M^2} \!\!\!\! d M'^2 
\left. \frac{d \sigma^\D_{3\Pom}}{d M'^2 \, dt} \right|_{t=0} \;.
\eeq
Generalizing this expression to multi-pomeron exchanges, including proper normalization and including coherence effect via the nuclear form factor $F_A(t_{min})$ \cite{Capella:1999kv}, we finally obtain
\beq
\label{eq:Kappa}
\kappa_A(Y,y_M,b) = \frac{ 4\pi A T_A(b)}{\sigma^{tot}_{\vp p} (Y)} \,  \int_{M_{min}^2}^{M^2} \!\!\!\! d M'^2 
\left. \frac{d \sigma^\D_{\vp p}}{d M'^2 \, dt} \right|_{t=0} \, F_A^2(t_{min}) \;.
\eeq
Sometimes $\kappa_A(Y,y_M,b)$ is called an effective rescattering cross section. Note that the total cross section in the denominator depends on the full rapidity distance, i.e. the center-of-mass collision energy.

Then, the total cross section at a given impact parameter from the summation of all fan diagrams is given by \cite{Schwimmer:1975bv}
\beq
\sigma^{Sch}_{\vp A} (Y,b) = \frac{e^{\Delta Y} g_{\Pom \vp} g_{\Pom A}(b)}{\kappa_A(Y,Y,b)+1} \;,
\eeq
where the integration limit in $\kappa_A$ goes up to a maximal mass, which is usually chosen to be constrained by $\xP^{max} = 0.1$. The total cross section in the Schwimmer model is given in terms of DIS variables in eqs.~(\ref{eq:SchwimmerModel}) and (\ref{eq:ShadowingFunction}). The total inelastic cross section \cite{Bondarenko:2000uv,Boreskov:2005ee} is
\beq
\sigma^{inel, \, Sch}_{\vp A} (Y,b) = \frac{e^{\Delta Y} g_{\Pom \vp} g_{\Pom A}(b)}{2 \kappa_A(Y,Y,b)+1} \;,
\eeq
and, finally, the cross section of single diffraction of the incoming hadron into a hadronic system with mass, $y=\ln \left( M^2 \big/ s_0\right)$, larger than $y_M$, $ y \geq y_M$, is given by
\beq
\sigma^{\D, \, Sch}_{\vp A}(Y,y_M,b) =  \frac{e^{\Delta Y} g_{\Pom \vp} g_{\Pom A}(b) \left(  \kappa_A(Y,Y,b) - \kappa_A(Y,y_M,b) \right)}{ \left[2 \kappa_A(Y,Y,b) - \kappa_A(Y,y_M,b) +1 \right] \left[ \kappa_A(Y,Y,b) + 1\right]}  \;.
\eeq
Then, taking the derivative, we find the cross section for single diffractive production of a system with mass $M^2$ as
\beq
\frac{d \sigma^{\D, \, Sch}_{\vp A}}{d y_M} &=& \int d^2b \, \frac{g_{\Pom A}^2(b) \, g_{\Pom \vp} \, G_{3\Pom} \, e^{\Delta (Y+y_M)}}{\left[2 \kappa_A(Y,Y,b) - \kappa_A(Y,y_M,b) +1 \right]^2} \\ 
&=&  \int d^2b \, \frac{A T_A(b) \, \sigma_\Pom(Y) }{\left[2 \kappa_A(Y,Y,b) - \kappa_A(Y,y_M,b) +1 \right]^2} \, \left[ \frac{d}{dy_M} \kappa_A(Y,y_M,b) \right] \;,
\eeq
which can be written as
\beq
\frac{d \sigma^{\D, \, Sch}_{\vp A}}{d M^2} = \left. \frac{d \sigma^\D_{\vp p}}{dM^2 dt} \right|_{t=0}  \int d^2b  \,\frac{4 \pi A^2 T^2_A(b) }{\left[2 \kappa_A(Y,Y,b) - \kappa_A(Y,y_M,b) +1 \right]^2} \;.
\eeq
Going to DIS variables, we define the Bjorken-$x$ related to the collision energy, or the full rapidity distance, $Y \rightarrow x = Q^2 \big/ (s + Q^2)$ and the mass of the produced system to $y_M \rightarrow \xP = \left( M^2 + Q^2\right) \big/ \left( s + Q^2\right) = x/\beta $. The final formula is given by eq.~(\ref{eq:CoherentDiffraction}).

The fan diagram summation presented above was derived in the "soft" limit, i.e. assuming simple pomeron exchanges. The generalization to full $\vp p$ cross sections in eq.~(\ref{eq:Kappa}) is an {\it ansatz} that is known to be valid for the second-order rescattering contribution to the total cross section, see eq.~(\ref{eq:MultipleScattering}). This prescription respects QCD scaling violations, and thus we extend it to the full analysis of hard reactions off nuclei.

\end{document}